\documentstyle[twoside,fleqn,npb,epsf]{article}
\newcommand{\an}{\overline{\alpha}_0}
\newcommand{\as}{\alpha_s}
\newcommand{\asmz}{\alpha_s(M_Z)}

\newcommand{\gev}{\,{\rm GeV}}

\newcommand{\ee}{\mbox{$e^+e^-$~}}
\newcommand{\mean}[1]{\left< #1 \right>}
\newcommand{\fmean}{\mean{F}}
\newcommand{\fpert}{\fmean^{\rm pert}}
\newcommand{\fpow}{\fmean^{\rm pow}}

\newcommand{\mi}{\mu_{\scriptscriptstyle I}}
\newcommand{\mr}{\mu_{\scriptscriptstyle R}}
\newcommand{\anmi}{\overline{\alpha}_0(\mi=2\gev)}
\newcommand{\order}{{\cal O}}
\newcommand{\rbthm}{\rule[-2ex]{0ex}{5ex}}

\newcommand{\rbtnpbr}{\rule[-1.2ex]{0ex}{3.6ex}}

\newcommand{\eprint}[2]{{hep-#1/}#2.}

\newcommand{\ejrnls}[4]{{#1} #2 (#3) #4;}
\newcommand{\jrnl}[4]{{#1} {\bf #2} (#3) #4.}
\newcommand{\jrnls}[4]{{#1} {\bf #2} (#3) #4;}

\newcommand{\JHEP}{\bf JHEP}
\newcommand{\JPG}{{\em J.~Phys.\ }{\bf G}}

\newcommand{\NPB}{{\em Nucl.~Phys.\ }{\bf B}}

\newcommand{\PLB}{{\em Phys.~Lett.\ }{\bf B}}

\title{Event Shapes and Power Corrections in $ep$ DIS}

\author{K.\ Rabbertz\address{I.\ Physikalisches Institut,
    RWTH Aachen, D-52056 Aachen, Germany\\
    (On behalf of the H1 Collaboration)}}

\begin{document}

\begin{abstract}
  Deep-inelastic $ep$ scattering data, taken with the H1 detector at HERA, are
  used to study the event shape variables thrust, jet broadening, jet mass,
  $C$~parameter, and two kinds of differential two-jet rates over a large
  range of ``relevant energy'' $Q$ between $7\gev$ and $100\gev$.
  
  The $Q$ dependence of the mean values is fit to second order calculations of
  perturbative QCD applying power law corrections proportional to $1/Q^p$ to
  account for hadronization effects. The concept of these power corrections is
  tested by a systematic investigation in terms of a non-perturbative
  parameter $\overline{\alpha}_{p-1}$ and the strong coupling constant.
\end{abstract}

\maketitle

\section{INTRODUCTION}

Due to asymptotic freedom perturbative QCD is a powerful theoretical
tool for the investigation of strong interactions at high energies
$Q$. Given, however, the precision of today's measurements over a
large range of energy, deviations diminishing proportional to $1/Q^p$
are observed.  Depending on the sensitivity of the observable on
long-distance, non-perturbative effects, these power corrections may
be sizable.  Less inclusive quantities than e.g.\ the total cross
section of $e^+e^- \rightarrow {\rm hadrons}$, where $p=4$, exhibit
much larger corrections; in case of event shapes they are typically
proportional to $1/Q$.

One possibility to account for the non-perturbative formation of
hadrons out of quarks and gluons is to invoke phenomenological models
implemented in Monte Carlo generators.  As a drawback, one has to deal
with several models, each again comprising a large number of tunable
parameters.  Alternatively, analytical formulae either inspired by a
simple tube model~\cite{Feynman} or by a recent approach initiated by
Dokshitzer and Webber~\cite{pc:DWform1} may be employed. A first study
of event shapes and analytically parameterized power corrections in
$ep$ deep-inelastic scattering (DIS) was published
in~\cite{H1:Shapes}.  The results presented here are based on the
substantially improved and extended analysis in~\cite{KR:Diss}.

\section{EVENT SHAPE VARIABLES}

The event shapes investigated are two kinds of 1-thrust, $\tau$ and
$\tau_C$, the jet broadening $B$, the jet mass $\rho$, the $C$
parameter, and two kinds of differential two-jet rates, $y_{fJ}$ and
$y_{k_t}$, with $y$ denoting the transition value of an event from
$(2+1) \rightarrow (1+1)$ jets. All of these infrared and collinear
safe variables are defined in the Breit frame of reference. Except for
the two jet observables $y_{fJ}$ and $y_{k_t}$, they exploit the
property of this frame to maximally separate the current quark jet
from the proton remnant by evaluating the current hemisphere alone.
They are normalized to the total momentum resp.\ the total energy in
the current hemisphere; the precise definitions may be found
in~\cite{KR:Diss,HUM:DIS98}.  In case of the $y$ variables, the
employed jet algorithms, i.e.\ the factorizable JADE and the $k_t$
algorithm~\cite{def:jets}, implicitly isolate the proton remnant.

The energy scale relevant in DIS is given by the four-momentum
transfer $q$ of the scattered lepton as $Q := \sqrt{-q^2}$ and ranges
from $7\gev$ up to $100\gev$ in this study.  Note that a second
quantity, e.g.\ the Bj{\o}rken scaling variable \mbox{$x :=
  -q^2/2P\cdot q$} with $P$ being the four-momentum of the incoming
proton, is necessary to fix the DIS kinematics. Here, $x$ varies
between $\approx 10^{-3}$ and $\approx 0.5\,$.

\section{POWER CORRECTIONS}

According to~\cite{pc:DWform1} the energy dependence of any event
shape mean $\fmean$ can be written as sum of a perturbative part
$\fpert$ and a power suppressed contribution $\fpow$. Identifying for
simplicity the renormalization scale $\mr$ with $Q$, the perturbative
part, evaluated to $\order(\as^2)$ using DISENT~\cite{NLO:DISENT}, is
given by
\begin{equation}
  \label{eqn:Fpert}
  \fpert = c_{1,F}(x)\,\as(Q) + c_{2,F}(x)\,\as^2(Q)\,.
\end{equation}
Note that in contrast to \ee annihilation parton density functions are
necessary to describe the initial state in $ep$ DIS\@.  Apart from
other quantities the coefficients therefore explicitly depend on~$x$.

Anticipating power corrections proportional to $1/Q$ and $1/Q^2$ in
the case of $\mean{y_{k_t}}$, the simple ansatz of $\fpow = \lambda/Q$
resp.\ $\fpow = \mu/Q^2$ is tested. However, fits of $\lambda$ or
$\mu$ alone are not able to describe the data satisfactorily. Fitting
also $\asmz$ mostly leads to unacceptable values of $\asmz$ owing to
strong correlations. Triggered by the $x$ dependence of $\fpert$, one
could assume that $\fpow$ should be a function of $x$ as well. Yet,
for corrections with power $p=1$ this is not
expected~\cite{prc:Dasgupta}.

This calls for the more elaborate approach of~\cite{pc:DWform1},
referring infrared divergences in the perturbative expansion to the
power corrections.  Assuming the existence of a well-behaved effective
strong coupling even at scales below the infrared matching scale $\mi$
of $\approx 2\gev$, one can deduce the relative size of the
corrections ${\cal P}/Q$ to the means of different event shapes at the
expense of a new non-perturbative parameter $\an$:
\begin{eqnarray}
  \label{eqn:calP}
  \nonumber {\cal P} & = &
  a_F {\cal M}^\prime \frac{16\mi}{3\pi} \bigg[ \an(\mi) - \as(Q)\\
  && - \frac{23}{6\pi}\left( \ln\frac{Q}{\mi} + 1.45\right) \as^2(Q) \bigg]\,,
\end{eqnarray}
where $a_F$ is a calculable $F$ dependent coefficient.  To resolve
ambiguities a refinement to two-loop level leads to another global
factor, the {\em Milan factor} ${\cal M}$, where ${\cal M}^\prime =
2{\cal M}/\pi \approx 1.14$~\cite{pc:Milan}.

\begin{figure}[htb]
  \centering
  \mbox{\epsfxsize=0.225\textwidth\epsffile{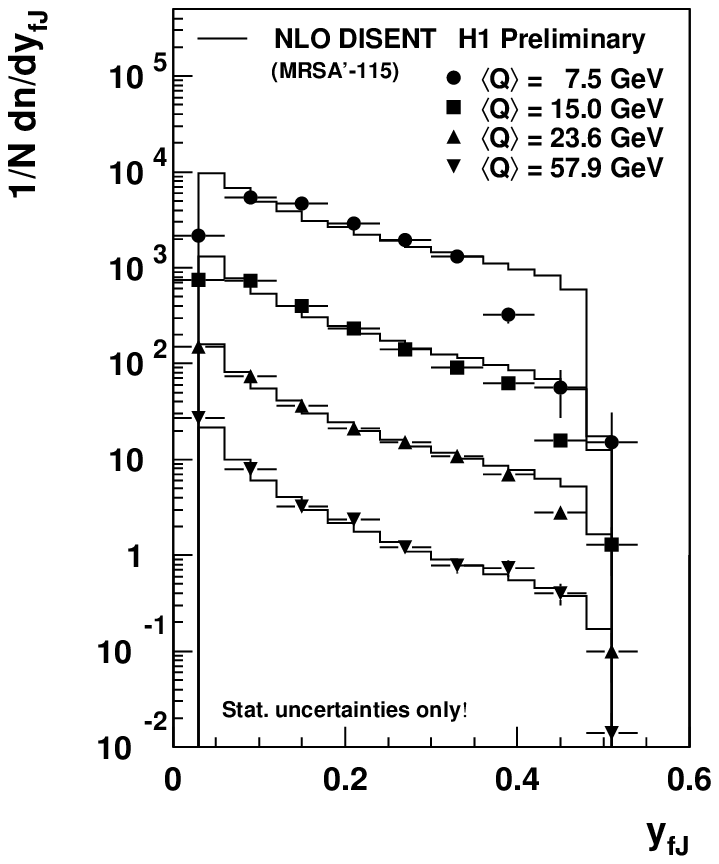}%
    \epsfxsize=0.225\textwidth\epsffile{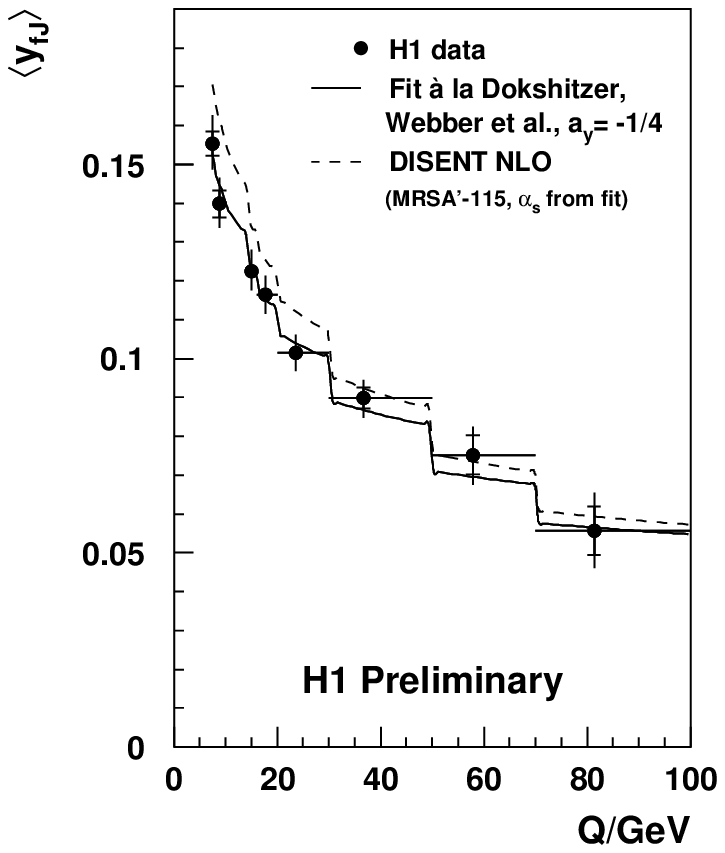}}
  \vspace*{-0.75cm}
  \caption{Left: Norm.\ diff.\ distr.\ of $y_{fJ}$ corrected for detector
    effects. H1 data (stat.\ uncert.\ only) are compared with NLO
    calculations.  Right: Corr.\ means of $y_{fJ}$ (stat.\ and tot.\ 
    uncert.)  as a function of $Q$. The full line corresponds to a
    power correction fit with $a_{y_{fJ}} = -1/4$.}
  \label{fig:fJADE}
  \vspace*{-0.75cm}
\end{figure}

As an example, fig.~\ref{fig:fJADE} shows the unfolded differential
distributions and a power correction fit for $y_{fJ}$, however, with
$a_{y_{fJ}} = -1/4$.

\begin{figure}[htb]
  \centering \epsfxsize=0.375\textwidth \epsffile{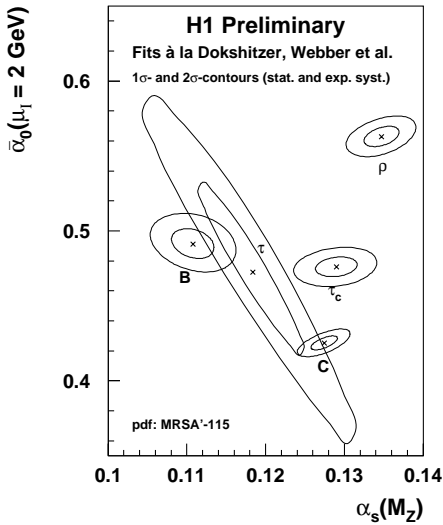}
  \vspace*{-0.75cm}
  \caption{Results of two-parameter power correction fits for the means
    of $\tau$, $B$, $\tau_C$, $\rho$, and $C$ in the form of $1\sigma$-
    and $2\sigma$-contours in the $(\as,\an)$-plane including stat.\ 
    and exp.\ syst.\ uncertainties.}
  \label{fig:ellipses}
  \vspace*{-0.75cm}
\end{figure}

The results of the two-parameter fits to this approach are presented
in table~\ref{tab:2parDWfit} and fig.~\ref{fig:ellipses}.  The
universal non-perturbative parameter $\anmi$ is confirmed to be
$\approx 0.5 \pm 20\%$ for the means of $\tau$, $B$, $\tau_C$, $\rho$,
and $C$, which are sizably affected by hadronization.  The spread in
$\asmz$, however, is uncomfortably large.  In contrast to earlier
analyses of $B$, a reasonable fit is obtained if the revised
coefficient $a_B^\prime$~\cite{pc:Brevisited} is included.

\begin{table*}[hbt]
  \caption{Results of two-parameter fits of $\anmi$ and $\asmz$
    according to the power correction approach initiated by
    Dokshitzer and Webber for the event shape means.
    The uncertainties are statistical
    and total systematic except for $y_{fJ}$ where a coefficient
    $a_{y_{fJ}}$ different from $1$ is anticipated.
    $\kappa$~denotes the correlation coefficient between the two
    fit parameters.}
  \label{tab:2parDWfit}
  \begin{tabular*}{\textwidth}{c@{\extracolsep\fill}ccccc}
    \hline
    \multicolumn{6}{c}{\rbthm H1 Preliminary}\\\hline
    $\rbthm\fmean$ & $a_F$ & $\anmi$ & $\asmz$ & $\chi^2/n$ &
    $\kappa/\%$ \\\hline
    $\mean{\tau}$ & $1$ &
    $0.480 \pm 0.028~^{+0.048}_{-0.062}$ &
    $0.1174 \pm 0.0030~^{+0.0097}_{-0.0081}$ &
    $0.5$ & $-97$ \rbtnpbr\\
    $\mean{B}$ & $1/2\cdot a_B^\prime$ &
    $0.491 \pm 0.005~^{+0.032}_{-0.036}$ &
    $0.1106 \pm 0.0012~^{+0.0060}_{-0.0057}$ &
    $0.7$ & $-58$ \rbtnpbr\\
    $\mean{\tau_C}$ & $1$ &
    $0.475 \pm 0.003~^{+0.044}_{-0.048}$ &
    $0.1284 \pm 0.0014~^{+0.0100}_{-0.0092}$ &
    $1.3$ & $+19$ \rbtnpbr\\
    $\mean{\rho}$ & $1/2$ &
    $0.561 \pm 0.004~^{+0.051}_{-0.058}$ &
    $0.1347 \pm 0.0015~^{+0.0111}_{-0.0100}$ &
    $1.2$ & $+7$ \rbtnpbr\\
    $\mean{C}$ & $3\pi/2$ &
    $0.425 \pm 0.002~^{+0.033}_{-0.039}$ &
    $0.1273 \pm 0.0009~^{+0.0104}_{-0.0093}$ &
    $0.9$ & $+63$ \rbtnpbr\\
    $\mean{y_{fJ}}$ & $1$ &
    $0.258 \pm 0.004$ &
    $0.104 \pm 0.002$ &
    $1.9$ & $-61$ \rbtnpbr\\\hline
    \multicolumn{5}{l}{$a_B^\prime \propto 1/\sqrt{\alpha_s} + {\rm const.}$
      as suggested by Dokshitzer, Marchesini and Salam~\cite{pc:Brevisited}.}
  \end{tabular*}
\end{table*}

Both $y$ observables exhibit small negative hadronization corrections.
For $y_{fJ}$ a power correction with $p=1$ and $a_{y_{fJ}}=1$ was
proposed in the conference proceedings~\cite{pc:WDIS95}, which has not
been reconsidered in recent publications.  From the emerging very low
values for both, $\anmi$ and $\asmz$, one can rule out $a_{y_{fJ}}=1$.
Turning the argument around, a reasonable fit can be achieved with
$a_{y_{fJ}}=-1/4$ while presupposing the validity of
eq.~\ref{eqn:calP}.  For $y_{k_t}$ fits with $p=1$ do not work
properly in accordance with the expectation of a power $p=2$.
Unfortunately, the coefficients $a_F$ are not yet calculable for that
case.

\section{SUMMARY AND CONCLUSIONS}

The improved and extended analysis of event shape means in $ep$ DIS
supports the concept of power corrections according to
eq.~\ref{eqn:calP}.  However, the uncomfortably large spread of values
gained for $\asmz$ leaves room for further progress.

Like in \ee annihilation, it would be very interesting to compare to
the power corrections obtained with event shape distributions.  To
that goal, resummed calculations are awaited for.

\section{ACKNOWLEDGEMENTS}
I would like to thank M.~Dasgupta and G.P.~Salam for enlightening discussions.

\end{document}